\documentclass{article}
\usepackage{INTERSPEECH2019,amsmath,graphicx}
\PassOptionsToPackage{hyphens}{url}\usepackage{hyperref}

\title{A New Approach to Accent Recognition and Conversion for Mandarin Chinese}
\name{Lin Ai$^1$, Shih-Ying Jeng$^2$, and Homayoon Beigi$^3$}

\address{
  $^{1,2}$Columbia University\\
  $^3$Recognition Technologies, Inc. and Columbia University}
\email{$^1$la2734@columbia.edu, $^2$sj2909@columbia.edu, and $^3$beigi@recotechnologies.com}

\begin{document}
%
\maketitle
\begin{abstract}
Two new approaches to accent classification and conversion are
presented and explored, respectively. The first topic is Chinese
accent classification/recognition. The second topic is the use of {\it
 encoder-decoder models} for end-to-end Chinese accent conversion,
where the classifier in the first topic is used for the training of
the accent converter {\it encoder-decoder model}. Experiments using
different features and model are performed for accent
recognition. These features include MFCCs and spectrograms. The
classifier models were TDNN and 1D-CNN. On the {\it MAGICDATA} dataset
with 5 classes of accents, the TDNN classifier trained on MFCC
features achieved a test accuracy of $54\%$ and a test F1 score of
$0.54$ while the 1D-CNN classifier trained on spectrograms achieve a
test accuracy of $62\%$ and a test F1 score of $0.62$. A prototype of
an end-to-end accent converter model is also presented. The converter
model comprises of an encoder and a decoder. The encoder model
converts an accented input into an accent-neutral form. The decoder
model converts an accent-neutral form to an accented form with the
specified accent assigned by the input accent label. The converter
prototype preserves the tone and foregoes the details in the output
audio. An {\it encoder-decoder} structure demonstrates the potential
of being an effective accent converter. A proposal for future
improvements is also presented to address the issue of lost details in
the decoder output.
\end{abstract}
\section{Introduction}
\label{sec:intro}
Accent variation is one of the most critical issues of the
state-of-the-art automatic speech recognition (ASR) systems,
especially for Mandarin Chinese. As a language with many dialects,
including Wu (spoken in Shanghai, Jiangsu and Zhejiang provinces) and
Yue (spoken in Cantonese areas such as Hong Kong and Guangdong),
Mandarin is spoken with significant variations, depending on speakers'
regional dwelling across the country. Therefore, it is very
challenging for any ASR system trained on standard Mandarin to perform
well while encountering speakers with varied accents across the
country. Adapting a sophisticated Chinese accent classifier or
recognition system could provide a strong improvement to the current
ASR systems.\\

In addition, accent conversion can also be a great solution to improve
ASR performance, in which a differently accented Chinese speech can
be converted to a standard Chinese dialect. Moreover, accent
conversion is of interest itself not only because it could possibly
improve ASR performance, but because it may be advantageous in many
other applications and use cases, such as second language learning.\\

Currently, much of the work done in the accent conversion domain is
limited to pairwise training and conversion, which requires a model to
be built between each pair of accents. This is a significant limitation,
given that there are so many possible accents, and it is absolutely not
feasible to train an additional model for each single pair of
accents. Furthermore, most of the current work and research focus only
on English, for example, different types of non-native English accents
versus native American or British dialects.\\

In this work, we propose and compare two types of Chinese accent
classifier models. One is a time delay neural network (TDNN) model
trained through transfer learning. The other one is a one-dimensional
(1D) convolutional neural network (CNN) model. We also present an
end-to-end Chinese accent conversion model, which is built using an
{\it encoder-decoder model} and one of our pre-trained accent classifiers.\\

The remainder of this paper is structured as
follows. Section~\ref{sec:literature} reviews some previous work on
accent conversion, voice conversion, and accent
recognition. Section~\ref{sec:methodology} introduces the detailed
methodology of our accent classifier models and the accent converter
model. Section~\ref{sec:res} describes the experiments we conducted,
in detail, and presents corresponding results. Finally,
Section~\ref{sec:conclusion_future} summarizes the conclusions of this
study, and discusses the potential work that we plan to carry out in
the future.\\

\section{Related Work}
\label{sec:literature}
As mentioned, most of the work done on accent conversion is limited to
pairwise training and conversion. Aryal et al. (2015)
\cite{Aryal2015ArticulatorybasedCO} train a deep neural network (DNN)
articulatory synthesizer for a non-native speaker, then map the
non-native articulatory space to a native speaker via Procrustes
transformations, and drive the trained DNN. They evaluate their model
through listening tests of intelligibility, voice identity, and
non-native accentedness. Bearman et al. (2017)
\cite{bearman2017accent} present a neural network model that learns
differences between a pair of accents and produces transformation
between the pair of accents using the extracted MFCC
features~\cite{r:beigi-sr-book-2011}. Their pairwise binary classifier
achieves an accuracy of $98.2\%$ between American English and Indian
accented English. Nonetheless, as they reveal in the paper, the
reconstructed waveforms are guttural and noisy, because MFCC features
may not always retain sufficient information for quality audio
reconstruction. Zhao et al. (2019) \cite{zhao2019foreign} use an
acoustic model trained on a native English speech corpus to extract
speaker-independent phonetic posteriorgrams (PPGs), and train a speech
synthesizer to map non-native speech PPGs into desired native spectral
features, which are then reconstructed into high-quality waveforms.\\

A similar domain that has been studied a lot is speaker voice
conversion. Mobin et al. (2016) \cite{mobin2016voice} apply CNN to
transform the voice of one speaker into another by manipulating not
only the pitch, but also the timbre~\cite{r:beigi-sr-book-2011}. They
also employ generative adversarial networks (GANs) to enhance their
generative model's performance. Mohammadi et al. (2014)
\cite{mohammadi2014voice} train a deep autoencoder to build
representations of short-term spectra of multiple speakers, which
enables voice conversion in a speaker-independent fashion.\\

As for accent detection, most work has been done on native and
non-native English accents. Jiao et al. (2016) \cite{jiao2016accent}
propose a combination of long-term and short-term training to tackle
both prosodic and articulation characteristics that differentiate
accents. DNNs are used for long-term statistical features training,
whereas recurrent neural networks (RNNs) are used for short-term
acoustic features training. They managed to achieve a classification
accuracy of $52.48\%$ over the $11$ accent classes. Sheng et al. (2017)
build a CNN model to classify $3$ different non-native English accents,
and achieve a classification test accuracy of $88.0\%$ over the $3$ accent
classes.  Hernandez et al. (2018) \cite{hernandez2018deep} train a
neural network to classify speech accents in video games, and achieve
a classification test accuracy of $71\%$ over $2$ accent classes.\\

Very little work has been done on Mandarin or other Chinese
dialects. Zheng et al. (2005) \cite{zheng2005accent} propose an
approach to combine accent detection and accent adapted model
selection for Chinese speech recognition. They build a Gaussian
mixture model (GMM) accent classifier with MFCC features, and achieve
an test accuracy of $86\%$ on the accented audio group. They then
apply MAP/MLLR to enhance acoustic adaptation and model selection, and
attain state-of-the-art acoustic modeling on Wu-accented Chinese
speech, reducing the character error rate by an absolute amount of
$1.0\%$ to $1.4\%$.\\

\section{Methodology}
\label{sec:methodology}
This section presents the methodology for the two main topics presented
here, namely, {\it accent recognition} and {\it accent conversion}.\\

\subsection{Accent Recognition}
\label{ssec:acc_reco}
Our proposed full accent converter model is composed of two parts: an
accent recognition model component, and an accent conversion
component. The accent conversion model training process is based on
the accent recognition model. Therefore, an accent recognition model
must be trained separately before training a complete end-to-end
accent converter model. The end-to-end accent converter model
structure is described in detail in Section~\ref{ssec:acc_conv}. This
section presents two different classifier model designs, using different
speech feature sets, TDNN classifier on MFCC features, and 1D-CNN
classifier on spectrogram features, respectively.\\

\subsubsection{TDNN Classifier on MFCC}
\label{sssec:tndd_mfcc_classifierl}
The first set of features are MFCCs, which have been widely used for
decades and usually produces state-of-the-art results in speaker
recognition~\cite{r:beigi-sr-book-2011}, speech recognition, and many
other related tasks in practice. Accent recognition is quite related
to the speaker recognition problem, in the sense that accent is an
important characteristic in distinguishing speakers. Since speaker
recognition~\cite{r:beigi-sr-book-2011} is a more complex and
better-studied area than accent recognition, it is reasonable to train
a speaker recognition model first and perform transfer learning to do
accent classification. Therefore, MFCC is selected for this experiment,
as it is generally used in speaker recognition tasks.\\

{\it x-vectors} \cite{snyder2018x} provide robust neural network
embeddings speaker recognition, and once combined with a customary
{\it Linear Discriminant Analysis (LDA)} and {\it Probabilistic Linear
  Discriminant Analysis (PLDA)}~\cite{r:beigi-sr-book-2011}, they
achieve superior performance on various speaker recognition evaluation
datasets. Therefore, training an {\it x-vector} model on Mandarin
corpus is the first step of this process. Using the {\it x-vectors} as
features for additional NN layers and a log softmax output layer, a
transfer learning, we build a transfer learning process and train an
accent classification model. The details of training process and model
architecture is described in Section~\ref{sssec:tdnn_mfcc_exp}.\\

\subsubsection{1D-CNN Classifier on Spectrogram}
\label{sssec:cnn_spec_classifierl}
The second set of feature used, was the spectrogram. Spectrograms have
demonstrate empirical effectiveness in accent detection and
recognition \cite{hernandez2018deep}. As a visual representation of
the spectrum of frequencies of signal for different time slices,
spectrograms resemble an images with one dimension representing
time. Therefore, image recognition techniques may be used directly on
spectrograms.  Convolutional Neural Networks (CNNs) have been
successfully used to perform machine learning on images
prolifically. Therefore, for the spectrogram features, we chose a CNN
as classifier.\\

The spectrogram input for one audio file is in 2 dimensional
format. Comparing this representation to images, it resembles
gray-scale images for which there is only a single color channel, or
the depth is 1 in the 3 dimensional representation. However, the
semantics of the width dimension is very different from gray-scale
images. The semantic of the width in spectrogram is time, which has a
special nature of being presented in sequence along time. With this in
mind, the CNN model chosen is 1D-CNN instead of 2D-CNN. While 2D-CNN
is commonly used and has proven success for regular images, the
semantic meaning of convolving spectrogram with 2D kernels which
crosses both different frequencies and time at the same time of the
convolution operation is unclear. In 2D images, the height and the
width dimension could be considered to be the same concept or in the
same domain, whereas in spectrogram it may not make sense to mix
frequency and time in the same kernel. Due to the nature of having a
time dimension, 1D-CNN is considered to be more suitable for machine
learning on spectrogram data in our design. It is also important to
note that Time Delay Neural Networks used in the previous section are
also a type of one dimensional convolutional neural network.  So, in
nature, the two architectures are not very different.  They just
operate on different features (MFCC vs Spectral).  The experimental
implementation may be found in Section~\ref{sssec:cnn_spec_exp}.\\

\subsection{Accent Conversion}
\label{ssec:acc_conv}
The accent converter model is an {\it encoder-decoder model}. The
encoder takes, as input, features of the original audio and converts them
to their {\it accent-neutral} representation\footnote{We
understand that every dialect has a specific accent associated with it.
By {\it accent-neutral}, we do not mean there is no
accent, but we simply imply that there is a standard accent
with possibly a majority of speakers, which may be used as the
reference accent.}, in the same feature space.  The decoder then take
the output of the encoder, which is the {\it accent-neutral} representation
of the input in the input feature space, together with an accent label
specifying the desired accent, and converts the encoded output into an
accented features with the specified accent. The input to the encoder,
the intermediate {\it accent-neutral} form (the output of the encoder), and
the output of the decoder are all in the same feature space. Namely,
the dimension of the encoder input, the encoder output, and the
decoder output are identical. There are two inputs to the accent
converter, which are the original audio's features, and the desired
accent label in one-hot format. There is one output from the accent
converter, which is the accent-converted audio's features. As an
accent conversion system that takes in an audio file and outputs
another audio file, preprocessing of extracting the features from
the audio file and postprocessing of converting the features back to
an audio file are necessary in addition to the converter model.\\

Different features could be used for the accent converter. One
requirement for such features is that they should be able to be
extracted from audio files (such as wav and mp3) and show also be
usable in reproducing an output audio file. Ideally, the chosen
features should help reduce the dimension/size of the data while
preserving sufficient information for successful accent recognition
and reconstruction of the audio file with an acceptable quality. In
our first prototype,we use spectral features. Other features such as
CELP~\cite{r-m:speex-manual} and combination of multiple features are
also worth considering.  CELP and other features related Linear
Predictive Coding (LPC)~\cite{r:beigi-sr-book-2011} have been used for
speech compression for decades and are prime candidates for usage in
this manner.  We will consider this in our future work (See
Section~\ref{sssec:alt_feat}).\\

The following two subsections describe the training process of the accent
converter and the inference/test workflow of the accent converter,
respectively.\\

\subsubsection{Training}
\label{sssec:acc_conv_training}
\begin{figure}[!htb]
\begin{minipage}[b]{1.0\linewidth}
  \centering
  \centerline{\includegraphics[width=8.5cm]{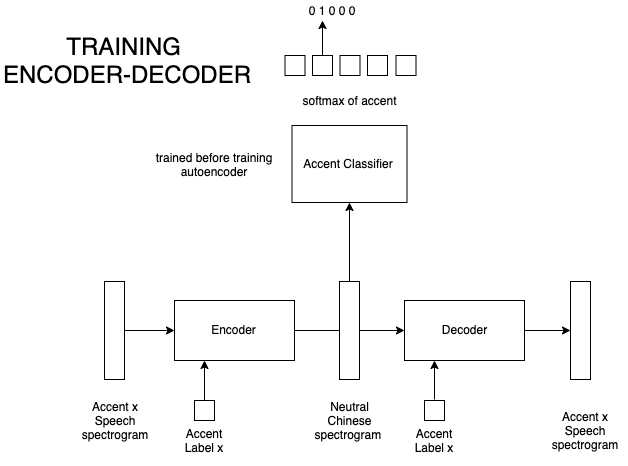}}
  \centerline{}\medskip
\end{minipage}
\caption{Converter training workflow}
\label{fig:converter_train}
\end{figure}

To train the converter to convert accented speech into {\it accent-neutral}
speech, an accent classifier is introduced. An accent classifier which
recognizes the accent class of speech is first trained using the
features that will be used in the accent converter. The class labels
are in one-hot format. After training the classifier, its weights will
be fixed and it could be used to assess the accent score for each
known accent in a speech in the feature space. Once the classifier is
ready, it is used as part of a trainer model. The trainer is the
{\it encoder-decoder model} with the intermediate output of the encoder
connected to the fixed weight pre-trained classifier. The high-level
structure of the trainer model is shown in
Fig.\ref{fig:converter_train}.\\

The trainer model has two inputs and two outputs.
\begin{itemize}
    \setlength\itemsep{-0.1em}
    \item Input 1: encoder input -- the original accented speech in the feature space
    \item Input 2: decoder input 2 -- the desired output accent label in one-hot format
    \item Output 1: classifier output -- the probability of the speech containing each accent as a vector
    \item Output 2: decoder output -- converted accented speech in the feature space 
\end{itemize}

The trainer model is composed of the encoder, the decoder, and the
classifier. The connective relations among these modules are as follows:
\begin{itemize}
    \item encoder output is connected to the classifier as the classifier input
    \item encoder output is connected to the decoder as the decoder input 1
\end{itemize}

The losses at both output branches, the classifier output and the
decoder output, are back-propagated through the model. The two losses
collectively guide the trainer model to learn. At the training time,
trainer input 1 is the original speech feature, the trainer input 2 is
the accent label of the original speech, the output 1 ground truth
label used is a uniform probability distribution, as a vector, and the
output 2 ground truth label is the original speech feature, identical
to model input 1. As an example, the output 1 ground truth label for
the {\it MAGICDATA} dataset (See Section~\ref{sssec:MAGICDATA}), with 5
accent classes, would be $<0.2, 0.2, 0.2, 0.2, 0.2>$. Given this
construction, with proper and sufficient training and in an ideal
scenario, the output of the encoder should eventually produce
{\it accent-neutral} speech in the feature space.\\

One potential drawback of this method is that there would never be
training pairs whose input accent is different from the converted
accent ground truth label. In the training the model is at best able
to reconstruct the original input after performing conversion. This is
a limit posted by the nature of the data, that it is not practical to
have the same person speak multiple different accents.\\

At the converter training time, preprocessing and postprocessing for
the conversion between input audio file and the speech features are
already taken care of as a preparation step for the training. The
trainer only deals with inputs and outputs in the feature space
(spectrogram in our experiment). At inference time, preprocessing
and postprocessing must be included to achieve an end-to-end
conversion system, as described in \ref{sssec:acc_conv_test}.\\

\subsubsection{Inference}
\label{sssec:acc_conv_test}
\begin{figure}[!htb]
\begin{minipage}[b]{1.0\linewidth}
  \centering
  \centerline{\includegraphics[width=8.5cm]{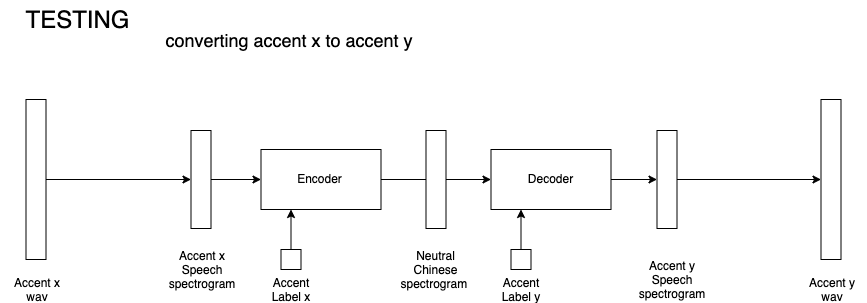}}
  \centerline{}\medskip
\end{minipage}
\caption{Converter inference workflow}
\label{fig:converter_inference}
\end{figure}

After the training process is completed via the trainer, the encoder
and decoder will ideally have proper weights for the accent conversion
task. The accent converter model is the combination of the trained
encoder and the trained decoder. The inference/test workflow of the
accent converter is shown in Fig.\ref{fig:converter_inference}.\\

As the encoder and the decoder are trained on features of the speech
instead of the original audio file, preprocessing of feature
extraction from the audio file and postprocessing for the purpose of
reconstruction from feature to audio are necessary components for
completing the system workflow, producing an end-to-end accent conversion.\\

The converter model has two inputs and one output.
\begin{itemize}
    \setlength\itemsep{-0.1em}
    \item Input 1: encoder input -- original accented speech in the feature space (after preprocessing)
    \item Input 2: decoder input 2 --  desired output accent label in one-hot format
    \item Output: decoder output -- converted accented speech in the feature space (before postprocessing) 
\end{itemize}

The accent converter model is composed of the encoder and the decoder
after they are trained using the trainer model. The connection
relation between the encoder and the decoder is as follows:
\begin{itemize}
    \item encoder output is connected to the decoder as decoder input 1
\end{itemize}

At the converter inference time, additional preprocessing and
postprocessing for conversion between input audio and the speech
features are added so that the system takes as input, an audio file (such
as wav or mp3) and a desired accent label (one-hot format) and produces
the accent-converted audio.\\

\section{Experiments and Results}
\label{sec:res}
This section describes the datasets used in our experiments, the
implementation details and results of the accent recognition models,
and the implementation details and results of the accent conversion
models.\\

\subsection{Data}
\label{ssec:data}
Two corpora used are {\it Aishell-2}~\cite{du2018aishell} and {\it
MAGICDATA}~\cite{MAGICDATA}.  Here, each dataset is described briefly.\\

\subsubsection{Aishell-2 Corpus}
\label{sssec:Aishell-2}
The {\it Aishell-2} \cite{du2018aishell} is a Chinese Mandarin speech corpus
published by Beijing Shell Technology Co., Ltd. The contents and
descriptions of the full corpus are as follows:
\begin{itemize}
    \setlength\itemsep{-0.1em}
    \item $1000$ hours of speech data (around $1$ million utterances)
    \item includes segmented transcripts
    \item $1991$ speakers ($845$ male and $1146$ female)
    \item provides speaker demographic information including age, gender, and accent region (north or south)
    \item recorded in indoor environments using high fidelity microphone and downsampled to $16kHz$
    \item manual transcription accuracy is above $95\%$
\end{itemize}
{\it Aishell-2} is by far the largest open-source Mandarin speech corpus and
it was used to train a speaker recognition model, which was used as a
pre-trained model to perform the transfer learning on accent
recognition.\\

One drawback of {\it Aishell-2} is that it labels accent region only in two
categories of north and south. Since there are many accents across the
country, dividing them purely by north and south is not desirable
grouping for our purposes. For example, the Shanghai accent of
Mandarin (Wu dialect spoken area) is quite different from the
Guangdong accent Mandarin (where Cantonese is also spoken), but they
are both labeled as one southern accent; whereas the Beijing accent
(usually considered as standard Mandarin) is labeled as northern even
though it shares much common with the Shanghai accent of
Mandarin. Therefore, labeling accents by province is much more
reasonable than simply tagging them northern or southern. This is
where the {\it MAGICDATA} corpus comes into place, where it provides more
fine-grained labels on accents, labeled by province.\\

\subsubsection{MAGICDATA Corpus}
\label{sssec:MAGICDATA}
\begin{figure}[!htb]
\begin{minipage}[b]{1.0\linewidth}
  \centering
  \centerline{\includegraphics[scale=0.55]{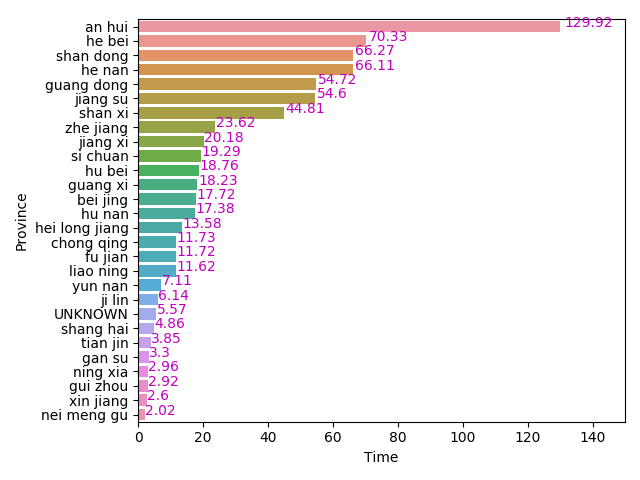}}
  \centerline{}\medskip
\end{minipage}
\caption{{\it MAGICDATA} training set data distribution, in terms of time (hours) versus province.}
\label{fig:MAGICDATA_dist}
\end{figure}

The {\it MAGICDATA} Mandarin Chinese Read-Speech Corpus~\cite{MAGICDATA} is
developed by MAGICDATA Technology Co., Ltd. The contents and
descriptions of the corpus are presented here:
\begin{itemize}
    \setlength\itemsep{-0.1em}
    \item $755$ hours of speech, mostly mobile recorded data
    \item includes segmented transcripts
    \item $1080$ speakers from different accent areas in China
    \item provides speaker demographic information including age, gender, and accent region (by province)
    \item sentence transcription accuracy higher than $98\%$
    \item recordings collected in quiet indoor environments
    \item speech data coding and speaker information file
    \item diversified domain of recording text, including interactive Q\&A, music search, SNS messages, home command and control
    \item Training set, validation set, and test set in a ratio of $51:1:2$
\end{itemize}
As mentioned in Section~\ref{sssec:Aishell-2}, {\it MAGICDATA}
provides fine-grained labels on speakers' accents by a province
label. The training set contains speakers from $28$ provinces, and the
test set portrays speakers from $8$ provinces. The data distribution
over the provinces is very unbalanced, as depicted in
Fig.\ref{fig:MAGICDATA_dist}. To balance the data distribution, we
focused on a subset of provinces, and grouped them into $5$ classes by
accent similarities and geographical proximity, as shown in table
\ref{table:MAGICDATA_classes}.\\

\begin{table}[!htb]
    \begin{center}
        \begin{tabular}{ |c | c |}
        \hline
        \textbf{class label} & \textbf{provinces}\\ \hline
        chuan & si chuan $\cup$ chong qing\\ \hline
        dongbei & ji lin $\cup$ liao ning $\cup$ hei long jiang\\ \hline
        guan & bei jing $\cup$ tian jin $\cup$ he bei\\ \hline
        wu & zhe jiang $\cup$ shang hai $\cup$ jiang su\\ \hline
        yue & guang dong $\cup$ guang xi\\ \hline
    \end{tabular}
    \end{center}
    \caption{{\it MAGICDATA} classes}
    \label{table:MAGICDATA_classes}
\end{table}

\subsubsection{Feature Extraction}
\label{ssec:feature_extraction}
For the training, two audio features were used: MFCC and spectrogram.\\

For MFCC features, $30$ cepstral coefficients are extracted with a
frame-length of $25ms$. Audios are sampled to $16kHz$. The resulting MFCC
features is with dimension $30$ per frame.~\cite{r:beigi-sr-book-2011}\\

Spectrogram features were extracted by using the {\it WORLD
 Vocoder}~\cite{morise2016world}, with a Fast Fourier transform
transform (FFT) size of $256$, and a frame period of $10ms$. The choice
of a relatively small FFT size was due to memory constraints. The audio was
sampled at $16kHz$. The resulting spectrogram features produce a
dimension of $(129, T)$, where $T$ is the audio length.\\

\subsection{Accent Recognition}
\label{ssec:res_acc_reco}
In this section two two accent classification models are introduced
and used in our experiments.\\

\subsubsection{TDNN on MFCC With Transfer Learning}
\label{sssec:tdnn_mfcc_exp}
As mentioned in Section~\ref{sssec:tndd_mfcc_classifierl}, an {\it
 x-vector} speaker recognition embedding is trained following the
same model structure as in~\cite{snyder2018x}. We followed the
Kaldi toolkit recipe for {\it VoxCeleb-v2} (VoxCeleb2) provided at
\url{https://github.com/kaldi-asr/kaldi/tree/master/egs/voxceleb/v2}. {\it
 Aishell-2} data was used to train the {\it x-vector} model because it is
by far the largest Mandarin speech corpus, containing $1991$
speakers with a balanced demographic distribution. Only $40\%$ of the
full corpus was used to reduce the training time. Utterances were
selected at random to prevent any unbalanced distribution. The details
about the data split are shown in table \ref{table:xvector_model_details}.\\

$3$-fold data augmentation was used following the approach
of~\cite{snyder2018x}, which randomly adds background speech (babble),
music, and noise, and applies reverberation to the original
recordings, and combines the original recordings, with two augmented
copies. MFCC features were extracted as described in
Section~\ref{ssec:feature_extraction}. The model training
configuration and train/validation accuracies are presented in table
\ref{table:xvector_model_details}.\\

LDA/PLDA~\cite{r-m:ioffe-2006} transformations with an output dimension of
$200$ were also trained to transform the {\it x-vectors} from their original
$512$ dimensions to lower dimensional space, more suitable for discriminating
the speaker class labels. A trial file of $44784$ pairs of
utterances was selected from the test set for scoring. The resulting EER
and DCF results are listed in table
\ref{table:xvector_model_details}. The model achieves an EER of
$3.4\%$, which is in tune with the {\it x-vector}
model trained on the VoxCeleb2 corpus~\cite{snyder2018x}.\\

\begin{table}[!htb]
    \begin{center}
        \begin{tabular}[width=5cm]{ | c | c | }
        \hline
        \textbf{Data Split} & \textbf{Training Configuration} \\ \hline
        Train set: $195944$ utterances & Number of epochs: $3$ \\ \hline
        Dev set: $27994$ utterances & Number of iterations: $80$ \\ \hline
        Test set: $55984$ utterances & Initial learning rate: $0.001$ \\ \hline
        \textbf{Training Results} & Momentum: $0.5$ \\ \hline
        Train accuracy: $98.3\%$ & Loss: Cross-entropy \\ \hline
        Validation accuracy: $97.9\%$ & Metric: Accuracy \\ \hline
        Trial file EER: $3.392\%$ & - \\ \hline
        minDCF($p=0.01$): $0.499$ & - \\ \hline
        minDCF($p=0.001$): $0.057$ & - \\ \hline
    \end{tabular}
    \end{center}
    \caption{{\it x-vector} data split, training configuration and model results}
    \label{table:xvector_model_details}
\end{table}

At this point, we performed transfer learning using the pre-trained
{\it x-vector} embedding. As mentioned in
Section~\ref{sssec:MAGICDATA}, {\it MAGICDATA} provides fine-grained
labels on accent areas by province, which is more suitable than {\it
 Aishell-2} for the accent classification task. Therefore, only {\it
 MAGICDATA} was used during the transfer learning process. The $5$  accent
classes are chuan, dongbei, guan, wu, and yue, as described in
Section~\ref{sssec:MAGICDATA}. The number of utterances used for
training and data split details are shown in
table~\ref{table:transfer_learning_details}.\\

\begin{table}[!htb]
    \begin{center}
        \begin{tabular}[width=5cm]{ | c | c | }
        \hline
        \textbf{Data Split} & \textbf{Training Configuration} \\ \hline
        Train set: $100135$ utterances & Number of epochs: $10$ \\ \hline
        Test set: $25030$ utterances & Number of iterations: $135$ \\ \hline
        \textbf{Training Results} & Initial learning rate: $0.05$ \\ \hline
        Train accuracy: $80\%$ & Momentum: $0.5$ \\ \hline
        Validation accuracy: $53\%$ & Loss: Cross-entropy \\ \hline
        Test accuracy: $54\%$ & Metric: Accuracy \\ \hline
        F1 score: $0.54$ & - \\ \hline
    \end{tabular}
    \end{center}
    \caption{Transfer learning accent recognition model data split, training configuration and results}
    \label{table:transfer_learning_details}
\end{table}

To perform transfer learning, $3$ fully connected layers with
the $relu$ activation were appended after the sixth TDNN layer of the
pre-trained {\it x-vector model}, and a log softmax output layer was added to
the end to map the network output to $5$ accent classes. Model layers
and their dimensions are shown in table
\ref{table:tdnn_model_structure}. During the training, the learning rate
of the first $7$ pre-trained layers, including the $6$ TDNN layers and $1$ stats
pooling layer, were set to $0$, and the initial learning rate of the
added transfer learning layers were set to $0.05$. The detailed training
configuration is listed in table
\ref{table:transfer_learning_details}.\\

\begin{table}[!htb]
    \begin{center}
        \begin{tabular}[width=5cm]{ | c | c | c | c |}
        \hline
        \textbf{Layer} & \textbf{Layer} & \textbf{Total} & \textbf{Input x Output}\\ 
                       & \textbf{Context} & \textbf{Context} &  \\ \hline
        input & - & - & $F \times 30$ \\ \hline
        tdnn1 & $[t-2, t+2]$ & $5$ & $150 \times 512$ \\ \hline
        tdnn2 & $\{t-2, t, t+2\}$ & $9$ & $1536 \times 512$ \\ \hline
        tdnn3 & $\{t-3, t, t+3\}$ & $15$ & $1536 \times 512$ \\ \hline
        tdnn4 & $\{t\}$ & $15$ & $512 \times 512$ \\ \hline
        tdnn5 & $\{t\}$ & $15$ & $512 \times 1500$ \\ \hline
        stats pooling & $[0, T)$ & $T$ & $1500T \times 3000$ \\ \hline
        tdnn6 & $\{0\}$ & $T$ & $3000 \times 512$ \\ \hline
        fc1* & $\{0\}$ & $T$ & $512 \times 256$ \\ \hline
        fc2* & $\{0\}$ & $T$ & $256 \times 128$ \\ \hline
        fc3* & $\{0\}$ & $T$ & $128 \times 64$ \\ \hline
        output* & $\{0\}$ & $T$ & $64 \times N$ \\ \hline
    \end{tabular}
    \end{center}
    \caption{TDNN Model Structure. Layers with * were appended layers during transfer learning. In the input layer, $F$ represents the number of frames in an input utterance. In the output layer, $N=5$ as there are $5$ accent classes.}
    \label{table:tdnn_model_structure}
\end{table}

\begin{figure}[!htb]
\begin{minipage}[b]{1.0\linewidth}
  \centering
  \centerline{\includegraphics[width=7.0cm]{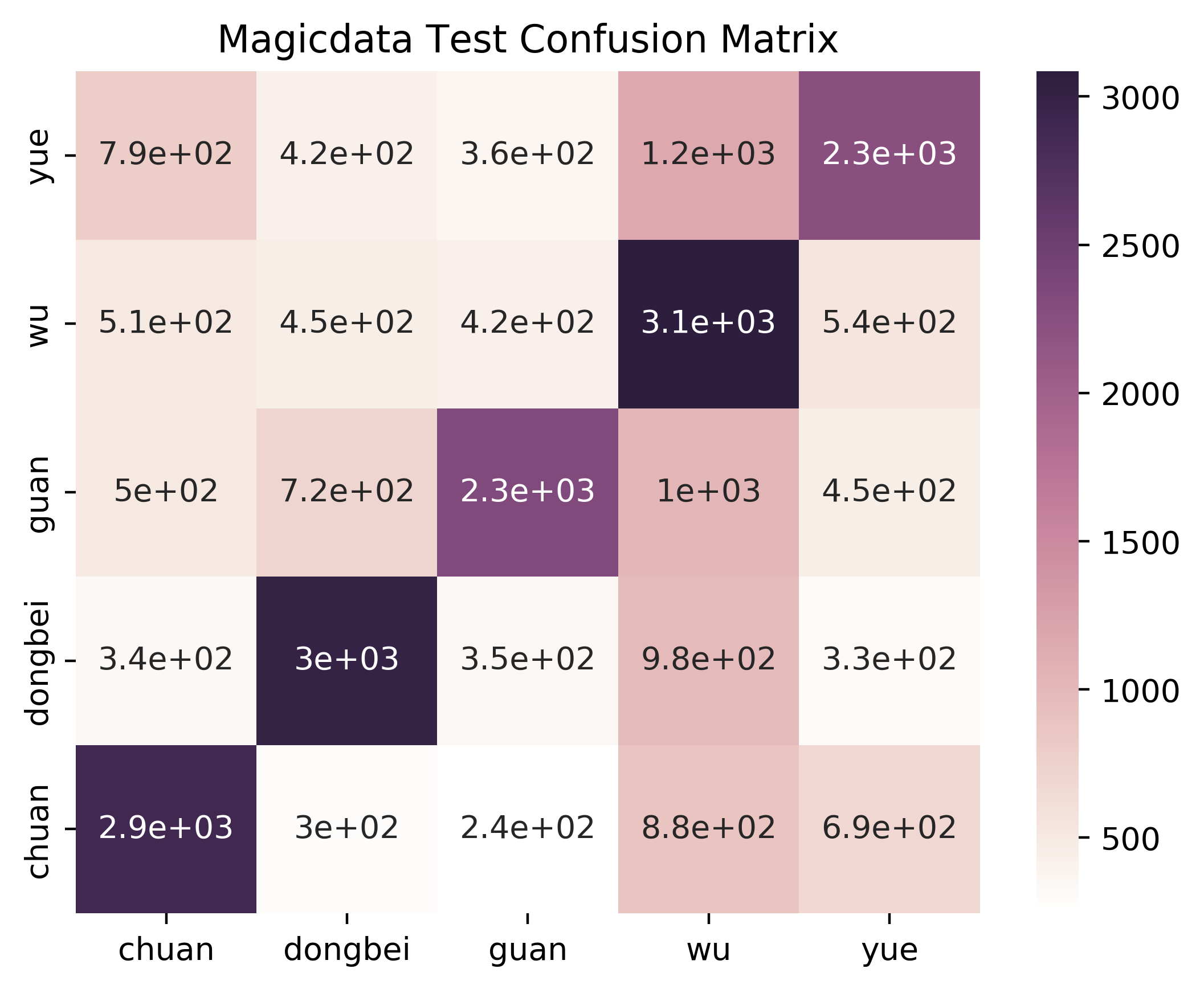}}
  \centerline{}\medskip
\end{minipage}
\caption{TDNN Classifier test confusion matrix}
\label{fig:tdnn_conf}
\end{figure}

The training results for the transfer learning process are also listed
in table~\ref{table:transfer_learning_details}. The model achieved a
test accuracy of $54\%$, and a classification F1 score of $0.54$. The
confusion matrix of this TDNN classifier on the test set for the $5$
accent classes is illustrated in figure \ref{fig:tdnn_conf}. From the
confusion matrix, it may be concluded that the TDNN classifier trained
through transfer learning can classify the {\it dongbei} accent and
the {\it wu} accent most easily, but it shows more trouble when
classifying the {\it guan} accent and the {\it yue} accent.\\

\subsubsection{1D-CNN on Spectrogram}
\label{sssec:cnn_spec_exp}
This section presents the implementation of the 1D-CNN classifier
described in \ref{sssec:cnn_spec_classifierl}.\\

\begin{figure}[!htb]
\begin{minipage}[b]{1.0\linewidth}
  \centering
  \centerline{\includegraphics[width=6.0cm]{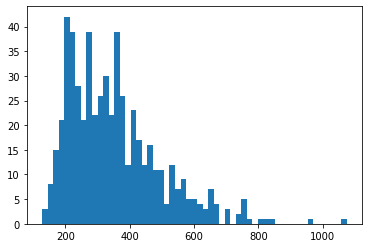}}
  \centerline{}\medskip
\end{minipage}
\caption{Time dimension distribution of spectrogram data}
\label{fig:spec_len_dist}
\end{figure}

To train a 1D-CNN model, the input to the 1D-CNN must be of a
predefined dimension and all input samples must have a predefined
dimension. In the spectrogram data extracted, as described in
\ref{ssec:feature_extraction}, the frequency axis is fixed while the
time dimension can vary depending on the original utterance length. To
unify the time dimension, we trimmed the long utterances and padded the
short ones. To determine the proper dimension for the time axis, the
distribution of time length, as shown in Fig.\ref{fig:spec_len_dist},
was taken into consideration. In our model, we set the time dimension
to $256$. For spectrogram with time longer than $256$, a random portion
of dimension $256$ was taken out and the exceeding part was trimmed. 
Spectrogram with shorter duration than $256$, we padded them with zeros.\\

The layers of 1D-CNN for the {\it MAGICDATA} is summarized in
\ref{table:1d_cnn_summary}.

\begin{table}[!htb]
    \begin{center}
        \begin{tabular}{ | c | c | c |}
        \hline
        \textbf{Layer Type} & \textbf{Output Shape} & \textbf{Params \#}\\ \hline
        Input & $[(None, 256, 129)]$ & $0$\\ \hline
        BatchNormalization & $(None, 256, 129)$ & $516$\\ \hline
        Conv1D & $(None, 247, 100)$ & $129100$ \\ \hline
        Conv1D & $(None, 238, 100)$ & $100100$ \\ \hline
        MaxPooling1D & $(None, 79, 100)$ & $0$ \\ \hline
        BatchNormalization & $(None, 79, 100)$ & $400$ \\ \hline
        Conv1D & $(None, 70, 160)$ & $160160$ \\ \hline
        Conv1D & $(None, 61, 160)$ & $256160$ \\ \hline
        GlobalAveragePooling & $(None, 160)$ & $0$ \\ \hline 
        Dropout & $(None, 160)$ & $0$ \\ \hline
        Dense Softmax Output & $(None, 5)$ & $805$ \\ \hline
    \end{tabular}
    \end{center}
    \caption{1D-CNN Model for {\it MAGICDATA} Summary}
    \label{table:1d_cnn_summary}
\end{table}


Batch normalization helps prevent the network training from
stagnating, due to the vanishing gradient problem and also provides a
some regularization. Dropout layer was also introduced to
regularize the training and to make the learning of the weights more
robust. Callbacks and early stopping were introduced to prevent
overfitting. The model training configuration is listed in table
\ref{table:1d_cnn_configuration}.\\

Experiments were carried out on both the {\it Aishell-2} and {\it
  MAGICDATA} datasets. The dimension of the spectrogram at every
timestamp was $129$, as specified before.\\

For the {\it Aishell-2} dataset with $2$ classes, the data split
details and training results are both illustrated in
table~\ref{table:1d_cnn_configuration}. Due to the nature of the
labels as described in~\ref{sssec:Aishell-2}, it is believed that the
results may not be conclusive enough, on the effectiveness of the
model. Therefore, experiment were carried out on the {\it MAGICDATA}.\\

For the {\it MAGICDATA} with $5$ classes, the data split details and training
results are both illustrated in table \ref{table:1d_cnn_configuration}
as well.\\

\begin{table*}[!htb]
    \begin{center}
        \begin{tabular}{ | c | c | c |}
        \hline
        \textbf{Training Configuration (for both datasets)} & \textbf{Aishell-2 Data Split} & \textbf{MAGICDATA Data Split}\\ \hline
        Number of epochs: $[20, 30]$ & Train set: $15672$ utterances (samples) & Train set: $12016$ utterances (samples)\\ \hline
        Loss function: Categorical cross-entropy & Test set: $4472$ utterances (samples) & Test set: $3004$ utterances (samples)\\ \hline
        Optimizer: Adam & Input shape: $(n, 255, 129)$ & Input shape: $(n, 256, 129)$ \\ \hline
        Metric: Accuracy & Output shape: $(n, 2)$ & Output shape: $(n, 5)$ \\ \hline
        - & \textbf{Aishell-2 Training Results} & \textbf{MAGICDATA Training Results} \\ \hline
        - & Train accuracy: $90\%$ & Train accuracy: $82\%$ \\ \hline
        - & Test accuracy: $87\%$ & Test accuracy: $62\%$ \\ \hline
    \end{tabular}
    \end{center}
    \caption{1D-CNN Model data split, training configuration and results}
    \label{table:1d_cnn_configuration}
\end{table*}

\begin{figure}[!htb]
\begin{minipage}[b]{1.0\linewidth}
  \centering
  \centerline{\includegraphics[width=7.0cm]{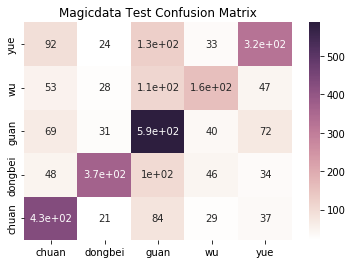}}
  \centerline{}\medskip
\end{minipage}
\caption{1D-CNN Classifier test confusion matrix}
\label{fig:1dcnn_conf}
\end{figure}

Fig.\ref{fig:1dcnn_conf} shows the confusion matrix of the 1D-CNN on
the test set, for the {\it MAGICDATA} dataset with $5$ classes. From
the confusion matrix, it can be concluded that the 1D-CNN classifier
performs best when classifying the {\it guan} accent, but has more
trouble classifying the {\it wu} accent.\\

\subsubsection{Classifier Comparison}
\label{ssec:classifier_comparison}
\begin{table}[!htb]
    \begin{center}
        \begin{tabular}{ | c | c | c | }
        \hline
        & \textbf{TDNN} & \textbf{1D-CNN} \\
        & \textbf{with MFCC} & \textbf{with Spectrogram} \\ \hline
        Train accuracy & $80\%$ & $\bf{82\%}$ \\ \hline
        Test accuracy & $54\%$ & $\bf{62\%}$ \\ \hline
        Best classified & {\it dongbei, wu} & {\it guan} \\ \hline
        Worst classified & {\it guan, yue} & {\it wu} \\ \hline
    \end{tabular}
    \end{center}
    \caption{Comparison of TDNN Classifier with MFCC and 1D-CNN Classifier with Spectrogram on {\it MAGICDATA}}
    \label{table:TDNN_CNN_comparison}
\end{table}

As illustrated in table \ref{table:TDNN_CNN_comparison}, 1D-CNN
classifier outperforms TDNN classifier, with a test accuracy of
$62\%$. This can be because the TDNN classifier is trained with MFCC
features, whereas the 1D-CNN classifier is trained using spectral
features. Spectral features contain different information compared with
MFCCs; specifically, spectrogram features contain pitch
information whereas MFCC features do not. Since Chinese is a tonal
language, pitch can be a key characteristics in differentiating
different accents. This difference in feature attributes could be the
essential reason why the 1D-CNN classifier outperforms the TDNN
classifier.\\

Another difference worth noticing is that TDNN classifier performs the
best on the {\it wu} accent, which 1D-CNN performs the worst on;
whereas 1D-CNN performs the best on guan accent, which TDNN classifier
performs the worst on. This could be caused by the similar issue
mentioned above, which is that spectrogram features contain pitch
information while MFCC features do not. From empirical experiences,
the {\it wu} accent and {\it guan} accent (usually considered standard
Mandarin) differ very little in tones, meaning both accents are
featured with standard Mandarin tones. It is the other
characteristics, such as differences in vowel and consonant
pronunciation, that distinguish the two accents. Therefore, pitch
information can be confounding when classifying the {\it wu} accent.\\

\subsection{Accent Conversion}
\label{ssec:res_acc_conv}
In the implementation of the accent conversion, spectrogram features
were used. The two classifiers trained, as in \ref{ssec:acc_reco}, use
MFCC features and spectrogram feature, respectively. As described in
\ref{ssec:acc_conv}, the feature used in the accent conversion must
have the property of being able to be converted to and from sampled
audio. MFCCs features do not provide the best reconstruction, and thus
are not as suitable for accent conversion. Therefore, in our accent
converter prototype, spectrogram feature were used. Other features
that may be reconstructed into audio form, such as Speex and
CELP~\cite{r-m:speex-manual}, are worth exploring in the future.\\

\subsubsection{Data Processing}
\label{ssec:data_proc_exp}
The first step of data processing was to unify the input spectrogram
dimension by trimming the long ones and padding the short ones, as
described earlier in \ref{sssec:cnn_spec_exp}. Since the classifier
model is part of the accent conversion trainer model, all the data
processing for the classifier and the converter must be identical. For
all the accent conversion experiments presented, the same data
processing steps were taken out and the classifier and the accent
converter were then trained on the same processed dataset.\\

The first few experiments were conducted on the raw spectrogram (after
unifying the dimension) without transforming the data. The classifier
performed similarly regardless of whether log and exponential
transformations and standardization were performed or not. On the
other hand, both the encoder and the decoder failed to learn on the
raw spectrogram (after unifying the dimension). The training suffered
from vanishing/exploding gradients.\\

The log-exponential spectrogram transformation presented in
\cite{hsu2016voice} was then implemented. The spectrogram was first
log-transformed, standardized, and then fed to the encoder. The output
from the decoder was destandardized, and then transformed
exponentially to retrieve the value of the scale before
transformation. The log-exponential transformation in the
implementation was based on the method in \cite{hsu2016voice}, with
the only difference of adding a small offset to prevent 0's in the
logarithms. With this transformation and the use of batch
normalization, the model trained properly, without experiencing any
exploding/vanishing gradients.\\

\subsubsection{Training}
\label{ssec:converter_traning_exp}
The initial attempt was to train the encoder and the decoder
separately, different from the training method described in
\ref{sssec:acc_conv_training}. To train the encoder and the decoder
separately, two separate trainers were built. The first trainer was
the encoder trainer, where the encoder was connected to the
fixed-weight classifier. The encoder trainer had one input and one
output as follows,
\begin{itemize}
    \setlength\itemsep{-0.1em}
    \item Input 1: encoder input -- original accented speech in the feature space
    \item Output 1: classifier output -- probability of the speech containing each accent, as a vector
\end{itemize}
The decoder trainer was the trained encoder connected to the decoder,
which was also the final converter model. The weights of the trained
encoder were fixed and only the decoder was trained. The decoder trainer
was also the converter, with two inputs and one output, as shown below.
\begin{itemize}
    \setlength\itemsep{-0.1em}
    \item Input 1: encoder input -- original accented speech in the feature space
    \item Input 2: decoder input 2 -- original accent label in one-hot format
    \item Output 1: decoder output -- converted accented speech in the feature space
\end{itemize}

With this technique of separating the training for the encoder and the
decoder, the converter did not perform well. In fact, the encoder
output produced parallel lines in the spectrogram, which was
undesirable, as it would not even preserve the content, let alone
being an {\it accent-neutral} representation of speech. The decoder
naturally failed because the decoder was based on the encoder's
output. The reason for the encoder learning a very lossy conversion
and outputting parallel lines could be that in this encoder trainer
model, the only output was the classifier output. In other words, the
loss incurred by the classifier output was the source of the overall
loss that solely guided the model's learning. The encoder would reach
a very low loss as long as the output could make the classifier output
a uniform classification prediction, without having to preserve the
speech information. To ensure that the encoder's output would preserve
the speech content and that it would be {\it accent-neutral}, the
encoder and the decoder should be trained together, as described in
\ref{sssec:acc_conv_training}. This way, both of the two output
losses (the classifier and the decoder output) would contribute to the
overall loss and collectively guide the model's learning. This can
prevent the issue previously encountered, when the encoder and the
decoder were trained separately. The training setting for the converter
trainer model on {\it MAGICDATA} dataset is provided here,
\begin{itemize}
    \setlength\itemsep{-0.1em}
    \item Loss for classifier output: categorical crossentropy
    \item Loss for decoder output: binary crossentropy
    \item Number of epochs: in range $[20, 30]$
    \item Batch size: $128$
    \item Train set size: $12016$
    \item Test set size: $3004$
\end{itemize}

\subsubsection{Latent Dimension}
\label{ssec:lat_dimension_exp}
Another experiment was performed on the encoder and the decoder model
structure. In this case, the intermediate result (encoder output) had
the same dimension as the encoder input and the decoder output, as it
is an {\it accent-neutral} representation of the speech in the same
feature space (s.t. it can be fed to the accent classifier). This
makes it very different from traditional autoencoder architectures,
where the introduction of a bottleneck latent dimension is key to
forcing a compressed knowledge representation of the original input
and does not just naively play the role of normalizing the input and
of passing the values through. We experimented with replacing both the
encoder model and the decoder model with an autoencoder architecture,
where a latent dimension was introduced. However, there did not appear
to be any significant improvement to warrant the benefit of this
architecture in our experiment. Eventually, the {\it encoder-decoder}
without any latent dimension was used.\\

\subsubsection{Converter Architecture}
\label{ssec:conv_arch}
The best accent converter model in our experiment was an {\it
  encoder-decoder model} trained on the {\it MAGICDATA} dataset. The
architecture of this model is shown in
Table~\ref{table:transformer_encoder_summary},
Table~\ref{table:transformer_decoder_summary_layer}, and
Table~\ref{table:transformer_decoder_summary_con}.
Table~\ref{table:transformer_encoder_summary} shows the encoder model
architecture.
\begin{table}[!htb]
    \begin{center}
        \begin{tabular}{ | c | c | c |}
        \hline
        \textbf{Layer Type} & \textbf{Output Shape} & \textbf{Params \#}\\ \hline
        Input & $[(None, 256, 129)]$ & $0$\\ \hline
        BatchNormalization & $(None, 256, 129)$ & $516$\\ \hline
        Conv1D & $(None, 256, 160)$ & $206560$ \\ \hline
        Conv1D & $(None, 256, 160)$ & $256160$ \\ \hline
        BatchNormalization & $(None, 256, 160)$ & $640$\\ \hline
        Conv1D & $(None, 256, 160)$ & $256160$ \\ \hline
        Conv1D & $(None, 256, 160)$ & $256160$ \\ \hline
        MaxPooling1D & $(None, 32, 160)$ & $0$ \\ \hline
        BatchNormalization & $(None, 32, 160)$ & $640$\\ \hline
        Dropout & $(None, 32, 160)$ & $0$ \\ \hline   
        Conv1D & $(None, 32, 100)$ & $160100$ \\ \hline
        Conv1D & $(None, 32, 100)$ & $100100$ \\ \hline
        Upsampling1D & $(None, 256, 100)$ & $0$ \\ \hline
        BatchNormalization & $(None, 256, 100)$ & $400$\\ \hline
        Conv1D & $(None, 256, 129)$ & $129129$ \\ \hline      
    \end{tabular}
    \end{center}
    \caption{Encoder Model for {\it MAGICDATA} Summary}
    \label{table:transformer_encoder_summary}
\end{table}
Table \ref{table:transformer_decoder_summary_layer} shows the decoder model layers and Table \ref{table:transformer_decoder_summary_con} shows the connection architecture.
\begin{table}[!htb]
    \begin{center}
        \begin{tabular}{ | c | c | c | }
        \hline
        \textbf{Layer Type} & \textbf{Output Shape} & \textbf{Params \#}\\ \hline
        Input1(Spectrogram) & $[(None, 256, 129)]$ & $0$\\ \hline
        Input2(Label) & $[(None, 5)]$ & $0$ \\ \hline
        Embedding & $(None, 5, 129)$ & $258$ \\ \hline
        Concatenante & $(None, 261, 129)$  & $0$ \\ \hline
        Conv1D-1 & $(None, 261, 160)$  & $206560$ \\ \hline
        Conv1D-2 & $(None, 261, 160)$  & $256160$ \\ \hline
        MaxPooling1D & $(None, 32, 160)$ & $0$ \\ \hline
        BatchNormalization-1 & $(None, 32, 160)$ & $640$ \\ \hline
        Conv1D-3 & $(None, 32, 100)$ & $160100$ \\ \hline
        Conv1D-4 & $(None, 32, 100)$ & $100100$ \\ \hline
        Dropout & $(None, 32, 100)$ & $0$ \\ \hline
        Upsampling1D & $(None, 256, 100)$ & $0$ \\ \hline
        BatchNormalization-2 & $(None, 256, 100)$ & $400$ \\ \hline
        Conv1D & $(None, 256, 129)$ & $129129$ \\ \hline      
    \end{tabular}
    \end{center}
    \caption{Decoder Model for {\it MAGICDATA} Layers}
    \label{table:transformer_decoder_summary_layer}
\end{table}

\begin{table}[!htb]
    \begin{center}
        \begin{tabular}{ | c | c | }
        \hline
        \textbf{Layer Type} & \textbf{Connected To}\\ \hline
        Input1(Spectrogram) & -\\ \hline
        Input2(Label) & -\\ \hline
        Embedding & Input2\\ \hline
        Concatenante & Input1 + Embedding\\ \hline
        Conv1D-1 & Concatenante\\ \hline
        Conv1D-2 & Conv1D-1 \\ \hline
        MaxPooling1D & Conv1D-2 \\ \hline
        BatchNormalization-1 & MaxPooling1D\\ \hline
        Conv1D-3 & BatchNormalization-1 \\ \hline
        Conv1D-4 & Conv1D-3\\ \hline
        Dropout & Conv1D-4 \\ \hline
        Upsampling1D & Dropout\\ \hline
        BatchNormalization-2 & Upsampling1D\\ \hline
        Conv1D & BatchNormalization-2 \\ \hline      
    \end{tabular}
    \end{center}
    \caption{Decoder Model for {\it MAGICDATA} Connections}
    \label{table:transformer_decoder_summary_con}
\end{table}

\subsubsection{Converter in Action}
\label{ssec:converter_demo}
Some
\href{http://www.recotechnologies.com/~beigi/ps/RTI-20200218-01/audio}{sample
  accent conversions} were run using the accented speech and its
corresponding accent class label as input. The ideal output of the
accent converter wold be the reconstruction of the input accented
speech. This experiment was performed with the {\it encoder-decoder
  converter} trained with $5$ accent classes of {\it MAGICDATA}.\\

\begin{figure}[!htb]
\begin{minipage}[b]{1.0\linewidth}
  \centering
  \centerline{\includegraphics[width=9.0cm]{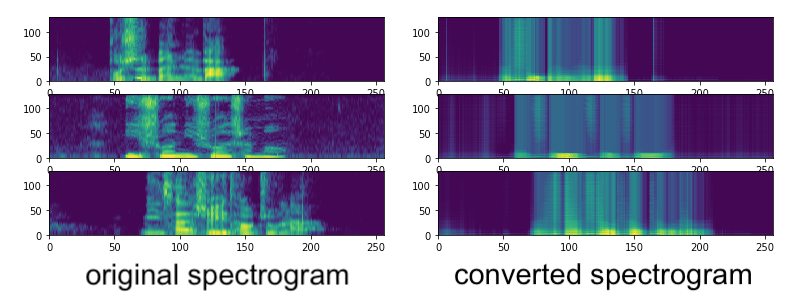}}
  \centerline{}\medskip
\end{minipage}
\caption{Original and converted spectrograms comparison}
\label{fig:spec_compare}
\end{figure}

Fig.\ref{fig:spec_compare} shows the comparison between the original
input spectrogram and the accent-converted spectrogram using the
original input's accent label, where the left side shows the original
spectrograms and the right side shows their corresponding converted
spectrograms (reconstructed via the converter).
As apparent in Fig.\ref{fig:spec_compare},  output spectrogram resembles
the input. The output preserving most of the lower frequencies while
losing details mostly in the higher frequencies.\\

\begin{figure}[!htb]
\begin{minipage}[b]{1.0\linewidth}
  \centering
  \centerline{\includegraphics[width=8.0cm]{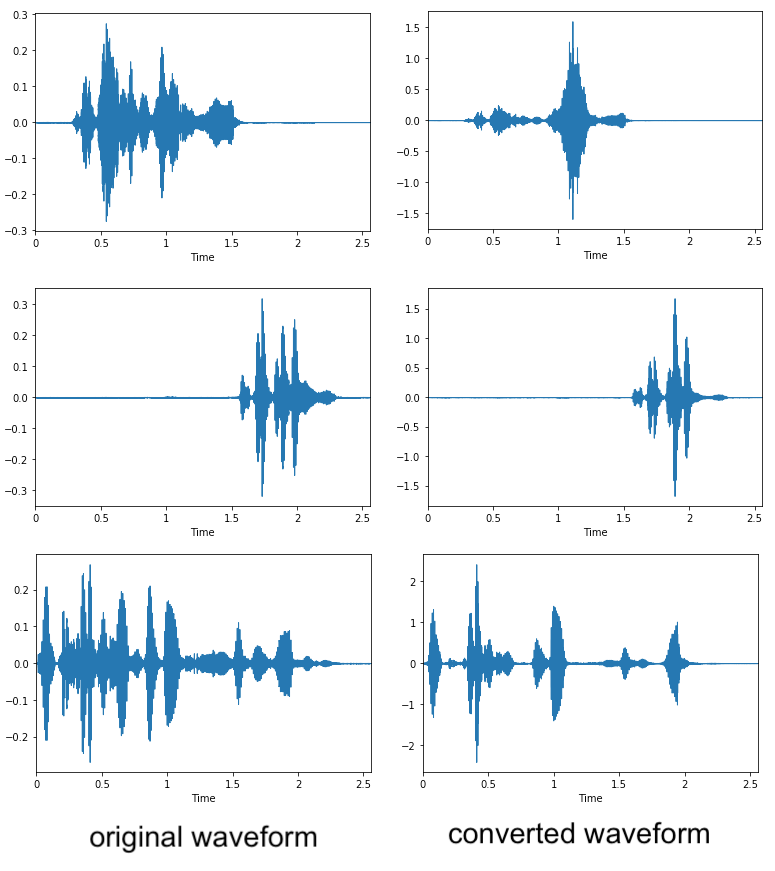}}
  \centerline{}\medskip
\end{minipage}
\caption{Original and converted waveforms comparison}
\label{fig:wav_compare}
\end{figure}

It is helpful to also look at the waveform of the speech input and
output.  Fig.\ref{fig:wav_compare} shows the comparison between the
original input waveform and the accent-converted waveform, using the
original input's accent label, where the left side shows the original
waveforms and the right side shows their corresponding converted
(reconstructed via the converter) waveforms. It is clear from
Fig.\ref{fig:wav_compare} that although the overall shape is similar,
the converted speech loses quite a bit of the detail, present in the
input.\\

The discovery from listening to the
\href{http://www.recotechnologies.com/~beigi/ps/RTI-20200218-01/audio}
{audio form of the sample accent conversions} is consistent with the visual
representation. The converted audio preserves the tone and intonation
of the input while the details are blurred.\\

A study on multi-target voice conversion without parallel data by
Chou, Yeh, Lee, and Lee \cite{chou2018multi} describes similar issues
of blurred output from the decoder and presents a solution. From their
insights, it is believed that the issue of losing details from the
decoder output may be addressed by the introduction of a cycle-GAN
model. We plan to pursue this approach in order to resolve this issue
of loss of details in the decoder output, a more detailed proposal of
future work to tackle this issue will be discussed in
\ref{ssec:future_work}.\\

\section{Conclusions and Future Work}
\label{sec:conclusion_future}
At this point some conclusions based on our new architecture and
approach are presented, followed by what will be pursued in some of
our future research.\\

\subsection{Conclusions}
\label{ssec:conclusion}
As shown in Section~\ref{ssec:classifier_comparison}, the 1D-CNN
classifier experiment outperforms the TDNN version. However, this is
most likely due to the use of spectral features in the 1D-CNN case,
which contain pitch information. Since Chinese is a tonal language,
pitch information can be a key characteristics in distinguishing
regional accents.  In addition, pitch, in any language defines the
major variations in accents.\\

As mentioned in Section~\ref{ssec:converter_demo}, converted speech
output from our converter model loses some details when compared with
the original spectrogram and waveform. By listening to the generated
audio, it is ascertained that the converted audio preserves the tones
and intonation of the original audio, but details are blurred. This is
a common issue with speech and audio generation, and needs further
improvement. One possible solution is described in
Section~\ref{ssec:future_work}. Being able to preserve tones and
intonation indicates that our converter model might perform better on
accents with distinctive tones and intonation.  This means that it
might produce better conversion results if the original accent and
desired accent have very different tones and intonation.\\

\subsection{Future Work}
\label{ssec:future_work}
In this section, we propose some of the experiments we may possibly
carry out in the future in order to improve our models.\\

\subsubsection{cycle-GAN for Decoder Output Refinement}
\label{sssec:cycle_gan}
As mentioned above, our converter model managed to preserve tones and
intonation during the conversion, but it blurred out the
details. Therefore, it is worth trying to tackle this issue using the
approach proposed by Chou, Yeh, Lee, and
Lee~\cite{chou2018multi}. This study on multi-target voice conversion
describes similar issues of blurred output from the decoder and
presents a solution. We believe that the issue of losing details from
the decoder output may be addressed by the introduction of a cycle-GAN
model.\\

\subsubsection{Transfer Learning on Spectrogram Features}
\label{sssec:transfer_spec}
Even though, currently, our 1D-CNN model outperforms the TDNN model
trained through transfer learning, we still believe that pre-trained
{\it x-vector} speaker recognition model might contain accent information,
and can be used for accent recognition. As discussed above, Chinese is
a tonal language, but MFCC features do not carry pitch
information. Therefore, one possible way to improve the TDNN model is
to combine pitch features with MFCC features, and/or use spectrogram
features during training.\\

As indicated by the results presented in
Table~\ref{table:TDNN_CNN_comparison}, the Spectrogram and MFCC
features seem to provide complementary results when it comes to
classifying the different accents.  Therefore, it seems quite
plausible that combining the MFCC with spectral features would
increase the accuracy of the underlying system.  In that regard, the
pitch may also be added to the set.\\

\subsubsection{Alternative Features}
\label{sssec:alt_feat}
One of the reasons why spectrogram features are of interest is that
they can be easily converted back to waveform, whereas there is
currently no simple and well-performing approach of generating
waveform with MFCC. From this aspect, we could also try to explore
some other features, such as CELP encoding, that can be easily
extracted/encoded and converted back/decoded to waveform. Such
features contain more information as well, which might possibly
improve our model performance.\\


\bibliographystyle{IEEEtran}
\bibliography{ms}
\end{document}